\documentclass[aps,prd,twocolumn,superscriptaddress,nofootinbib]{revtex4-2}

\usepackage{graphicx}
\usepackage{xspace}
\usepackage{xcolor}
\usepackage{hyperref}
\usepackage{booktabs}
\usepackage{amssymb}
\usepackage{siunitx}
\usepackage{amsmath}
\newcommand{\aligo}{Advanced LIGO\xspace}
\newcommand{\avirgo}{Advanced Virgo\xspace}
\newcommand{\gw}[1][]{gravitational wave#1\xspace}
\newcommand{\snr}[1][]{signal-to-noise ratio#1 (SNR#1)\renewcommand{\snr}[1][]{SNR##1\xspace}\xspace}
\newcommand{\psd}[1][]{power spectral density#1 (PSD#1)\renewcommand{\psd}[1][]{PSD##1\xspace}\xspace}
\newcommand{\imbh}[1][]{intermediate mass black hole#1 (IMBH#1)\renewcommand{\imbh}[1][]{IMBH##1\xspace}\xspace}
\newcommand{\cbc}[1][]{compact binary coalescence#1 (CBC#1)\renewcommand{\cbc}[1][]{CBC##1\xspace}\xspace}
\newcommand{\chisq}{$\chi^2$\xspace}
\newcommand{\sgd}[1][]{stochastic gradient descent#1\xspace}

\newcommand{\tensorflow}[0]{\textsc{Tensorflow}}

\newcommand{\msun}{M_\odot}

\begin{document}

\preprint{LIGO-P2100244}

\title{Using machine learning to auto-tune chi-squared tests for gravitational wave searches}

\author{Connor McIsaac}
\email[]{connor.mcisaac@port.ac.uk}
\affiliation{DISCnet Centre for Doctoral Training, University of Portsmouth, Institute of Cosmology and Gravitation, Portsmouth PO1 3FX, United Kingdom}
\author{Ian Harry}
\affiliation{University of Portsmouth, Institute of Cosmology and Gravitation, Portsmouth PO1 3FX, United Kingdom}

\date{25th May 2022}

\begin{abstract}

The sensitivity of \gw searches is reduced by the presence of non-Gaussian noise in the detector data. These non-Gaussianities often match well with the template waveforms used in matched filter searches, and require signal-consistency tests to distinguish them from astrophysical signals. However, empirically tuning these tests for maximum efficacy is time consuming and limits the complexity of these tests. In this work we demonstrate a framework to use machine-learning techniques to automatically tune signal-consistency tests. We implement a new \chisq signal-consistency test targeting the large population of noise found in searches for intermediate mass black hole binaries, training the new test using the framework set out in this paper. We find that this method effectively trains a complex model to down-weight the noise, while leaving the signal population relatively unaffected. This improves the sensitivity of the search by $\sim 11\%$ for signals with masses $> 300 \msun$. In the future this framework could be used to implement new tests in any of the commonly used matched-filter search algorithms, further improving the sensitivity of our searches.

\end{abstract}

\maketitle

\section{\label{sec:intro}Introduction}

Since the first detection of \gw[s] in 2015~\cite{GW150914}, the LIGO detectors have been upgraded multiple times, and the network of detectors now includes \aligo \cite{aLIGO}, \avirgo \cite{aVirgo} and Kagra \cite{Kagra}. The third observing run of the LIGO-Virgo-Kagra network \cite{LVK} has been completed and a whole suite of new \cbc signals have been observed \cite{GWTC2, GWTC2.1, GWTC3, 3OGC, 4OGC, IAS3}. Along with improvements in the detectors, search algorithms used to search for \cbc signals have also been improved by using information from the full network and introducing new methods for removing noise \cite{ExtPyCBC, gstlalO2, mbta}.

A set of modelled \cite{findchirp, ihope, pycbc, ExtPyCBC, gstlalearly, gstlalmid, gstlalO2, mbta, mbtaO3, spiir, IAS1, IAS2, IAS3} and unmodelled \cite{cwb} search algorithms have been used to observe \cbc[s] in the past. In this paper we will focus on modelled searches. Modelled searches construct a large bank of simulated signals (templates) with a variety of masses and spins in order to cover the targeted parameter space. These templates are then each used to perform a matched filter over the data for each detector. The matched filter \snr is then compared across the detector network in order to check for consistency across detectors.

The matched filter is the optimal solution when searching for known signals assuming that only Gaussian noise is present~\cite{findchirp}. However, in the presence of non-Gaussian noise transients, large \snr[s] can be produced where no signal is present~\cite{ihope}. Gravitational-wave detectors contain many such transients \cite{noise, o3detchar}, commonly referred to as ``glitches". Glitches can produce huge values of \snr while having little resemblance to the \cbc signals being searched for. It is therefore necessary to employ signal-consistency tests to remove as many of these glitches as possible in order to reduce the rate of false alarms. In this paper we will explore how to effectively develop and tune such a signal-consistency test to separate the signal and noise populations. We will apply this methodology to the PyCBC search \cite{ExtPyCBC} as a demonstration of its use. However, we strongly emphasize that this method will be applicable to any modelled search.

In the PyCBC search, multiple signal-consistency tests are employed. Firstly, the matched filter \snr is modified using two \chisq tests that compare the morphology of the signal in the data with that of the template \cite{BAchisq, SGchisq}, penalising any signals found to be inconsistent. Peaks in the re-weighted \snr time-series are then compared across the detector network, checking for consistency in the template parameters, as well as the relative time of arrival, amplitude and phase of the signals \cite{phase}. Each potential candidate event is then given a detection statistic in order to rank the likelihood that it is a real signal.

After these tests are performed, the remaining signals are shifted in time relative to each other to empirically measure a non-astrophysical background. The detection statistic values of the background signals can then be compared to the observed coincident signals in order to produce a false alarm rate for each observed signal \cite{pycbc}.

The existing signal consistency tests remove a large number of glitches from the single detector data, reducing the rate of background coincidences, and therefore reducing the false alarm rate of observed signals. However, a large number of glitches continue to be detected, particularly when searching with high mass templates \cite{pycbcimbh}, $M_{\mathrm{total}} \gtrsim 100 M_\odot$, where the signal may only be in the detector frequency band for a fraction of a second.

In this work we propose a method of creating and tuning new signal-consistency tests in order to separate signal and noise populations. We show that we can use machine-learning techniques such as \sgd to optimise these tests efficiently, and thus improve the sensitivity of our searches to \gw signals.

The use of machine-learning in \gw searches is an area where much work is being done. Several works have explored the use of neural networks to replace the matched filter statistic \cite{George:2016hay, George:2017pmj, Gabbard:2017lja, Gebhard:2019ldz, Yan:2021wml}, using convolution neural networks to predict the probability of a signal being present.

One advantage of convolution neural networks compared to a matched filter is the computational cost involved. This is particularly important for multi-messenger astronomy, where prompt detection of \gw signals could enable follow-up with electromagnetic observations. It has been shown that convolution neural networks could be an effective method for enabling such observations \cite{Wei:2020sfz, Wei:2020xrl, Wei:2020ztw, Schafer:2020kor, Krastev:2019koe, Krastev:2019koe}.

These works have shown that machine-learning can, in some cases, compete with the sensitivity of a matched filter search when applied to a single detector. However, such methods have not yet been demonstrated to be effective in large-scale searches for \cbc[s] covering a wide range of parameters. Current methods also do not produce additional information such as the amplitude and phase of the signal, used in the matched filter search to test triggers across detectors, they therefore lose sensitivity when compared to a matched filter search using a network of detectors \cite{Schafer:2021cml}.

In this work we choose to introduce a machine learning model within the current matched filter framework in order to utilise the statistical tests already available to us. We implement a new \chisq test using \sgd to train a set of tunable parameters within the model, optimising the test using noise triggers from a previous search along with a set of simulated signals. By implementing the model as a \chisq test it should remain rigorous in the case of unseen data, such as a new population of glitches.

We start by describing the general use of \chisq tests within \gw searches, and the tests currently used within the PyCBC search in Sec. \ref{sec:background}. In Sec. \ref{sec:chisq} we describe our proposed framework for training new \chisq tests using machine-learning methods. We then utilise this framework in Sec. \ref{sec:imbh} to train a new \chisq test for use in a search for \imbh[s]. In Sec. \ref{sec:results} we present the effect of this trained model on the \imbh search showing that it improves the sensitivity of the search, particularly at high masses, where the effect of non-Gaussian noise is most prominent.

\section{\label{sec:background}A Review of \chisq Tests in GW Searches}

We will begin by reviewing existing \chisq signal-consistency tests used in \gw searches.

In order to search for signals in strain data, a matched filter is used. Assuming the strain data takes the form of $s(t) = n(t) + h(t)$, where $n(t)$ is stationary Gaussian noise and $h(t)$ is a known signal, matched filtering is the optimal method for detecting the signal $h(t)$. The calculated \snr is analytically maximised over the amplitude and phase of the signal. The \snr is calculated as:
\begin{equation}
    \rho^2 = \frac{|(s|h)|^2}{(h|h)},
\end{equation}
where the inner product is:
\begin{equation}
    (a|b) = 4\int^{f_{\mathrm{high}}}_{f_{\mathrm{low}}}
    \frac{\Tilde{a}(f) \Tilde{b}^{\ast}(f)}{S_n(f)} df
\end{equation}
and $S_n(f)$ is the one-sided \psd of the noise. However, in the case of non-Gaussian noise, large peaks in the \snr time-series can be produced. Short bursts of non-Gaussian noise are often referred to as ``glitches", these can produce extremely large values of \snr with little similarity to the signal being searched for. In order to remove such triggers signal-consistency tests may be introduced to test the morphology of the trigger compared to that of the search template.

A \chisq test is one such test. A \chisq test is constructed by performing a matched filter with additional templates, $\hat{h}_{\bot}$, that are orthogonal to the search template such that $(h|\hat{h}_{\bot}) = 0$. Given well-behaved noise, and a signal which is an exact match to the search template, the matched filter \snr of the orthogonal template will follow a reduced \chisq distribution with 2 degrees of freedom~\cite{findchirp}. However, when there is non-Gaussian noise present, such as a glitch, the \snr will deviate from this distribution, taking a larger value. By examining triggers on the \snr-\chisq plane the signal and noise populations may then be separated.

After choosing a suitable template, $\hat{h}$, to be used for the \chisq test one first normalises it so that $(\hat{h}|\hat{h}) = 1$. The part of the signal orthogonal to the search template is then selected:
\begin{equation}\label{eq:ortho}
    \hat{h}_{\bot} = \frac{\hat{h} - (\hat{h}|h)h}{\sqrt{1 - (\hat{h}|h)^2}}.
\end{equation}
N such templates are created in this way and their \snr[s] are combined to produce the \chisq statistic.
\begin{equation}
    \chi^2_{r} = \frac{1}{2N}\sum^N_i\rho^2_i.
\end{equation}
In the case that the templates are orthogonal to one another this will produce a reduced \chisq distribution with $2N$ degrees of freedom. However, in general, orthogonality between the \chisq templates is not always enforced, in which case the statistic will follow a generalised \chisq distribution with increased variance.

The \chisq test also assumes that the signal in the data matches the search template. However, due to the discrete placement of templates within the parameter space of the search there will be some mismatch between these two signals. This mismatch means that the \chisq template will no longer be orthogonal to the signal and some of the signal's \snr will be included in the \chisq statistic, increasing the mean of the distribution, creating a non-central \chisq distribution. A similar effect will be caused if the \psd is miscalculated, or if the noise is non-stationary.

In general any N templates can be chosen to construct a \chisq test. However, this test will be most effective when templates are chosen that have some overlap with known non-Gaussian noise in the data, in particular, aiming to target noise which produces high \snr triggers for the targeted parameter space.

To separate the signal and noise populations a re-weighted \snr is then calculated that penalises triggers where the $\chi^2_{r}$ is larger than expected. This re-weighting takes the general form
\begin{equation}\label{eq:general_reweight}
    \hat{\rho} = f(\rho, \chi_r^2).
\end{equation}
This re-weighted \snr is then used to rank potential candidate events.

\subsection{\label{sec:existing} Existing \chisq Tests in the PyCBC search}

There are currently two \chisq signal-consistency tests employed within the PyCBC search. The first of these is the Allen \chisq test \cite{BAchisq}. This test divides the template into $p$ independent frequency bins, splitting the template such that each bin will contribute equally to the \snr. The \snr contribution for each of these sub-templates is calculated and compared to the expected value, calculating the \chisq statistic as
\begin{equation}
    \chi_r^2 = \frac{p}{2p - 2} \sum^{p}_{i=1} \left( \frac{\rho}{p} - \rho_{bin,i} \right)^2.
\end{equation}
This will take large values when a glitch is present in the data that does not share the same morphology as the search template. Specifically this test checks the distribution of power along the track of the \cbc signal. Although this test does not follow the exact form described in the previous section, it follows the same principle detecting excess power along the track of the signal. The re-weighted \snr \cite{chicurrent} is then calculated as
\begin{equation}\label{eq:BAreweight}
    \Tilde{\rho} = 
    \begin{cases}
    \rho, & \text{if $\chi_r^2 \leq 1$, } \\
    \rho \left[ \left(1 + (\chi_r^2)^3\right) /2 \right]^{-\frac{1}{6}}, & \text{if $\chi_r^2 > 1$.}
    \end{cases}
\end{equation}

By ranking the candidates based on this re-weighted \snr, a large number of noise triggers can be down-weighted. This test is particularly powerful for lower mass systems where the signals span a wide range of frequencies within the band of the detectors, allowing for a larger number of frequency bins to be used effectively. The number of frequency bins to be used is varied as a function of the search templates parameters \cite{chibins}. The number of frequency bins to use and the form of the \snr re-weighting are currently tuned empirically and have evolved over time \cite{chiold, chicurrent, chibins}.

The second \chisq test is the sine-Gaussian \chisq test \cite{SGchisq}. This works by performing a matched filter with $n$ sine-Gaussian signals, each being placed at frequencies higher than those expected from the search template. As these sine-Gaussian signals do not overlap with the search template, we can construct a \chisq test as the sum of their \snr[s]
\begin{equation}
    \chi^2_{r,sg} = \frac{1}{2n}\sum^n_i\rho^2_{sg,i}.
\end{equation}
This statistic tests if excess power is present above the final frequency of the search template. When excess power is present large values of $\chi^2_{r,sg}$ will be produced and the \snr is re-weighted again
\begin{equation}\label{eq:SGreweight}
    \Tilde{\rho}_{sg} = 
    \begin{cases}
    \Tilde{\rho}, & \text{if $\chi_r^2 \leq 4$}, \\
    \Tilde{\rho} \left(\chi^2_{r,sg} / 4 \right)^{-\frac{1}{2}}, & \text{if $\chi_r^2 > 4$}.
    \end{cases}
\end{equation}

The addition of this second test further reduces the rate of noise triggers due to glitches. This test has a significant impact for higher mass templates where there is a population of short duration glitches known as ``blips" \cite{blips}. A subset of these blips have power extending to high frequencies allowing this test to remove them successfully \cite{SGchisq}. However a large number of these glitches are not removed by this test \cite{pycbcimbh}.

Both of these tests have been tuned empirically by hand, choosing the number of frequency bins to be used and the placement of the sine-Gaussian signals, as well as the exact form of the re-weighting of the \snr. In the next section we propose an approach that allows us to create and tune new \chisq tests using a data-driven approach.

\section{\label{sec:chisq}Auto Tuning of a \chisq signal-consistency test}

We propose a framework in which we create new \chisq tests and empirically tune them based on a set of training data. To achieve this we take a set of noise triggers from a previous search along with a set of simulated signals and use \sgd to tune the parameters of our model.

In order to optimise the parameters of our chosen model we first must define a loss function, this is the quantity that we aim to minimise during the training process. The loss function is a function of the triggers \snr, $\rho$, and its \snr re-weighted by the new \chisq test described in Sec. \ref{sec:imbh}, $\hat{\rho}$. For this work we choose to define a separate loss function for noise triggers and simulated signals. The loss functions used in the case of noise triggers is,

\begin{equation}
    L_{n}(\hat{\rho}, \rho) = 
    \begin{cases}
    \hat{\rho} - 4, & \text{if $\hat{\rho} > 4$} \\
    0, & \text{if $\hat{\rho} \leq 4$}
    \end{cases}
\end{equation}

which penalises any cases where the re-weighted \snr is greater than 4. Below this threshold the PyCBC search currently discards all triggers so it is unnecessary to reduce it any further.

The loss function used for simulated signals is:

\begin{equation}
    L_{inj}(\hat{\rho}, \rho) = \rho - \hat{\rho}
\end{equation}

This penalises the case where the $\chi_r^2$ value is large and the \snr is reduced. This contribution to the loss will also allow us to train a function that re-weights the \snr as in Eq. \ref{eq:general_reweight} in order to create a greater separation between the signal and noise populations in the \snr-\chisq plane.

To update the parameters of the model we must then calculate the gradients with respect to the loss function. This is done using backpropogation after calculating the loss function using a set of training data. The \chisq model and matched filter are implemented in \tensorflow \cite{tensorflow2015-whitepaper, tensorflow_developers_2021_5177374}, allowing the gradients to be tracked through the calculation and the parameters updated.

Stochastic gradient descent has been used widely in the field of deep learning to optimise extremely complex models \cite{sgd, DeepLearning}, this framework therefore allows us to produce highly complex transformations while allowing us to effectively tune them to the detector data at a reasonable computational cost.

In the next section we will describe one such model and the data used to train the model.

\section{\label{sec:imbh}A \chisq test for intermediate mass black hole searches}

To demonstrate the training scheme described in the previous section we will attempt to train a \chisq test that improves the separation between glitches and signals when searching for \imbh[s], which we consider as signals with $M_{\rm tot}> 100 M_{\odot}$. In this mass range the Allen \chisq test described in the Sec. \ref{sec:existing} has a limited effect due to the systems merging at low frequencies and covering a relatively small frequency range within the detector bandwidth. The sine-Gaussian \chisq test is successful in removing a population of glitches that affect templates within this mass range, however, many glitches remain that do not have significant high frequency power~\cite{pycbcimbh}.

In this section we will define a transformation using the framework outlined in the previous sections using data from a previous \imbh search in order to train the \chisq test and improve the sensitivity of the search.

\subsection{\label{sec:model}Creating \chisq templates}

Existing \chisq tests effectively test the distribution of power of a candidate event along the track of the signal in the time-frequency plane and test for excess power at high-frequencies. We aim to test for excess power in a frequency band similar to those of the search templates, aiming to cover areas of the time-frequency plane not currently covered by existing tests.

To achieve this we transform the search template itself, shifting the template in time and frequency. The optimal values for these time and frequency shifts will depend on the template being used and the noise present in the data. Time shifted templates have previously been used in the PyGRB search \cite{Harry:2010fr, Williamson:2014wma} to create a \chisq signal-consistency test using fixed values for the time shifts. We propose a model that allows different time and frequency shifts for each template, tuning these values based on the current data.

The time and frequency shifts are selected using a dense neural network. The template is first turned into an input by sampling it between 12 Hz and 256 Hz with a sample width of 0.1 Hz, we then take the absolute value of the template and normalise it so that the mean of the input is one. This input is then passed to a dense neural network with two output values between -1 and 1. The first of these values is used as the time shift after being scaled by the maximum allowed time shift $\Delta t_{\mathrm{max}}$, similarly, the second value is used as the frequency shift after being scaled by the maximum allowed frequency shift $\Delta f_{\mathrm{max}}$.

The dense neural network consists of 11 dense layers, using the Rectified Linear Unit (\textsc{ReLu}) activation functions for hidden layers and the hyperbolic tangent function for the output layer. In order to produce multiple time/frequency shift pairs we train multiple networks with this configuration. However, to speed up the training of this model the first 6 dense layers are shared between each network. The sizes of the dense layers are listed in Table. \ref{tab:architecture}.

\begin{table}[t]
    \centering
    \begin{tabular}{l c | r}
         Layer & & Output Size \\
         \hline
         Input & & 2440 \\
         Dense + \textsc{ReLu} & * & 128 \\
         Dense + \textsc{ReLu} & * & 128 \\
         Dense + \textsc{ReLu} & * & 64 \\
         Dense + \textsc{ReLu} & * & 64 \\
         Dense + \textsc{ReLu} & * & 32 \\
         Dense + \textsc{ReLu} & * & 32 \\
         Dense + \textsc{ReLu} & & 16 \\
         Dense + \textsc{ReLu} & & 16 \\
         Dense + \textsc{ReLu} & & 8 \\
         Dense + \textsc{ReLu} & & 8 \\
         Dense + tanh & & 2 \\
    \end{tabular}
    \caption{This table details the architecture of the neural network used to calculate time and frequency shifts. The input is the absolute value of the search template sampled at 0.1 Hz between 12 Hz and 256 Hz. Dense layers transform their input using a matrix multiplication followed by the addition of a bias vector, the output is then passed to the activation function listed. The hyperbolic tangent activation function of the final layer ensures the output is in the range $[-1, 1]$. The two outputs are multiplied by $\Delta t_{\mathrm{max}}$ and $\Delta f_{\mathrm{max}}$ respectively to generate the time and frequency shifts. 4 such networks are generated, the layers marked with an asterisk (*) share their weights between the 4 networks.}
    \label{tab:architecture}
\end{table}

After calculating these shifts they are applied to the template before using Eq. \ref{eq:ortho} to generate our \chisq templates.

In addition to this model we must define a function to re-weight the \snr with the $\chi^2_r$ value. We create a parameterised model that can reproduce the re-weighting in Eq. \ref{eq:BAreweight} and \ref{eq:SGreweight}.
\begin{equation}\label{eq:reweight}
    \hat{\rho} = 
    \begin{cases}
    \rho, & \text{if $\chi_r^2 \leq \sigma$}, \\
    \rho \left(\left(\delta + (\chi^2_{r} / \sigma)^{\beta} \right) / \left(\delta + 1 \right)\right)^{-\frac{1}{\alpha}}, & \text{if $\chi_r^2 > \sigma$}.
    \end{cases}
\end{equation}
Here $\sigma$, $\alpha$, $\beta$ and $\delta$ are parameters that can also be tuned to increase the effectiveness of the test. This re-weighting leaves any signals with $\chi_r^2$ less than the threshold, $\sigma$, unchanged. At large $\chi_r^2$ values $\alpha$ and $\beta$ determine how quickly the re-weighted \snr decreases with the $\chi_r^2$ value, while $\beta$ and $\delta$ effect the transition between these two regimes.

\subsection{Data}

In order to most effectively train the model we target glitches that are missed by previous signal-consistency tests. We achieve this by performing a search using the setup described in \cite{pycbcimbh} covering $~ 45$ days of data from the first half of the third observing run. The data used in this search is available from GWOSC \cite{LOSC, GWOSC}. From this search we then select a set of noise triggers with $\Tilde{\rho}_{sg} > 6$ and $6 \leq \rho \leq 64$. These noise triggers may have been down-weighted by existing signal-consistency tests, but have not been down-weighted enough to remove them from the analysis. In order to avoid including real signals we remove any triggers that are coincident across multiple detectors. This will also remove a number of noise triggers that have formed coincident triggers, however, enough triggers remain to create a substantial training set. These triggers are then clustered over a window of 1 second and the triggers with the largest $\Tilde{\rho}_{sg}$ in that window are kept. We record the times of the triggers and the parameters of the template that produced them.

We also select a set of simulated signals to include during training. From the list of simulated signals that were recovered by the search, with false-alarm rates smaller than 1 per year, we select a set using the same constraints as the noise triggers. For these triggers we record the parameters of the simulated signal, as well as the template that recovered the signal. By using the template that recovered the signal in the search we are including the effect of template mismatch within the training scheme, this allows the \snr re-weighting step in Eq. \ref{eq:reweight} to be tuned to account for this contribution. The simulated signals used in this analysis include effects from precession and higher-order modes that are not present in the template bank, by including these effects in the analysis we can train the model to avoid identifying these effects as noise, maintaining our sensitivity to these signals.

For each sample, the strain is loaded at 2048Hz, for samples containing simulated signals the signal is then added. The strain data is high-pass filtered at 12Hz and PSDs are calculated using 512 seconds of data around the time of the sample, following the same procedure as the PyCBC search. 64 seconds of data around the trigger is then selected, ensuring the trigger is not within the first 32 seconds or the last 16 seconds to allow time for the inspiral and ringdown of the search templates.

The search templates are generated using the \textsc{SEOBNRv4\_ROM} \cite{SEOBNRv4_1, SEOBNRv4_2} waveform model, and are generated at the same length and sample frequency as the strain data. The simulated signals are generated using the \textsc{NRSur7DQ4} \cite{NRSur7DQ4}, \textsc{SEOBNRv4} \cite{SEOBNRv4_1, SEOBNRv4_2} and \textsc{SEOBNRv4HM} \cite{SEOBNRv4HM_1, SEOBNRv4HM_2} waveform models.

In order to ensure the noise and signal samples have similar importance during training we select an equal number of each. Additionally, to ensure that the parameters in Eq. \ref{eq:reweight} are trained to separate the noise and signal populations across a range of \snr[s] we bin our samples by \snr and select an equal number of samples from the noise and signal models for each bin. The boundaries of these \snr bins are 6, 8, 10, 14, 18, 24 and 64. In each of these \snr bins we draw 1200 noise samples and signal samples from those remaining after filtering has been applied.

We set aside 10\% of all samples to be used as a validation set to monitor the performance of the model on unseen data, all other samples are used to train the model. This gives us a total of 12960 samples for training and 1440 samples for validation.

\subsection{Training}

The training of the model is performed in batches, each batch contains 32 samples from the training set described in the previous section. For each sample the peak \snr within 0.1 seconds of the trigger time is calculated. The search template is then generated and passed to the transformation described in Sec. \ref{sec:model}. The resulting templates are then normalised and the orthogonal templates calculated using Eq. \ref{eq:ortho}. These templates are then used to calculate $\chi_r^2$ at the time of the peak \snr and the re-weighted \snr is calculated. This value can then be passed to the loss function in order to calculate the loss values for the batch.

Once the losses have been calculated, backpropogation is used to obtain the gradients of the loss function with respect to the trainable weights of the network described in Sec. \ref{sec:model}, as well as the trainable parameters in Eq. \ref{eq:reweight}. Stochastic gradient descent is then used to apply small changes to the variables based on the calculated gradients. In order to speed up training and improve performance for sparse gradients we use Nesterov momentum \cite{Nesterov1983AMF} when calculating the parameter updates.

Before training the full model we perform a pre-training phase where only the parameters of the \snr re-weighting in Eq. \ref{eq:reweight} are trained. This step is faster than the training of the full model and by performing this step first the training of the \chisq model is more effective early in the training. After this step we proceed with the main training step, training the parameters of the model described in Sec. \ref{sec:model} and the \snr re-weighting in Eq. \ref{eq:reweight} at the same time.

The main training step is repeated until all samples in the training set have been analysed 25 times, taking a total of $\sim 24$ hours using 8 CPU cores. During training the learning rate determines how quickly parameters are changed in response to the calculated gradients. In order to improve convergence late in the training stage we employ learning rate decay. After each full cycle of the training set the learning rate is multiplied by a factor of 0.9, allowing the model to make smaller adjustments late in training.

\subsection{Trained Model}

\begin{figure}[t]
    \centering
    \includegraphics[width=0.85\linewidth]{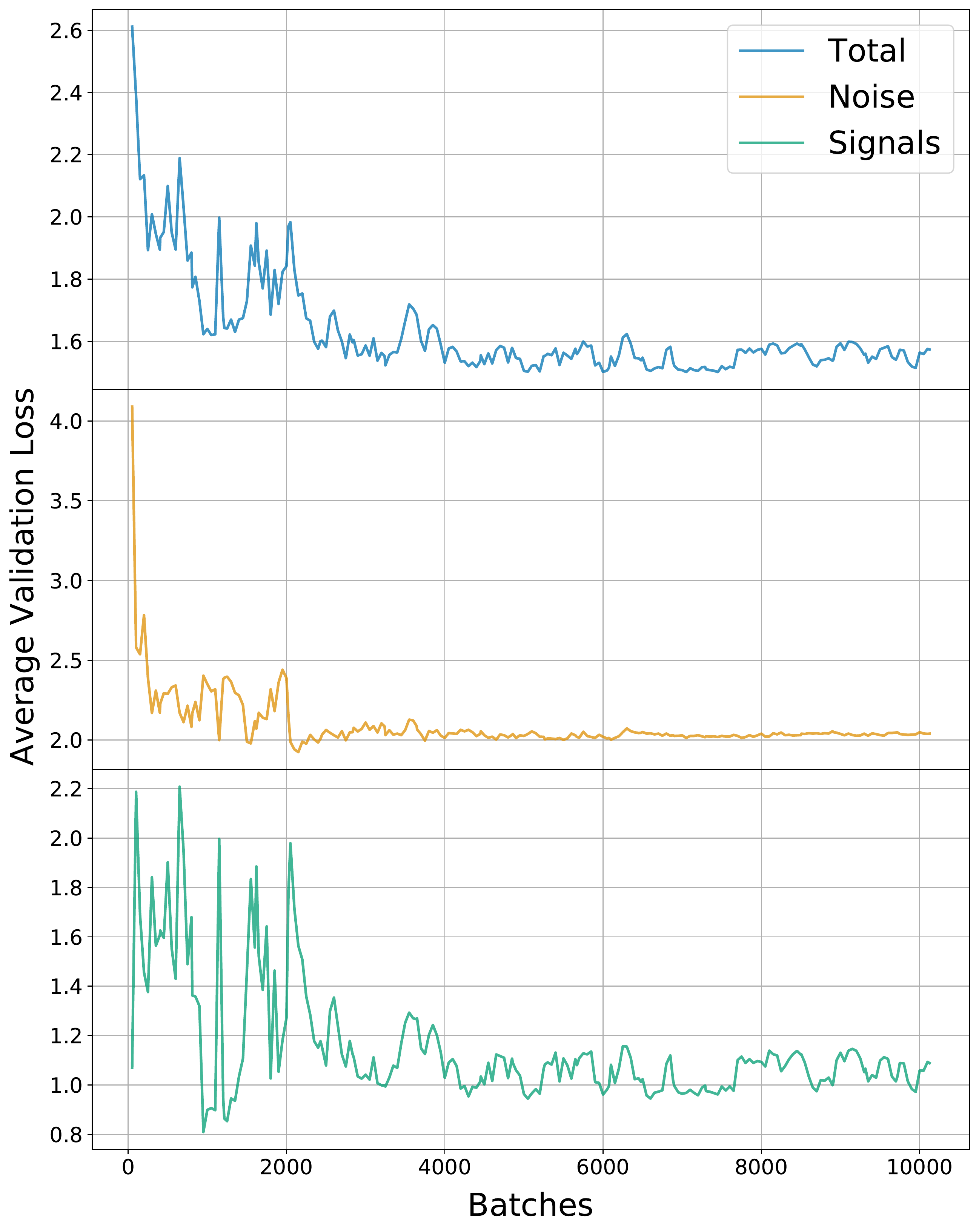}
    \caption{The average loss calculated using the unseen test data as it changes with the number of training batches used. The average losses contributed by noise samples and signal samples are also plotted. This shows a large improvement in the loss contributed by noise samples, while the loss from signals remains reasonably steady.}
    \label{fig:loss}
\end{figure}

\begin{figure}[t]
    \centering
    \includegraphics[width=0.85\linewidth]{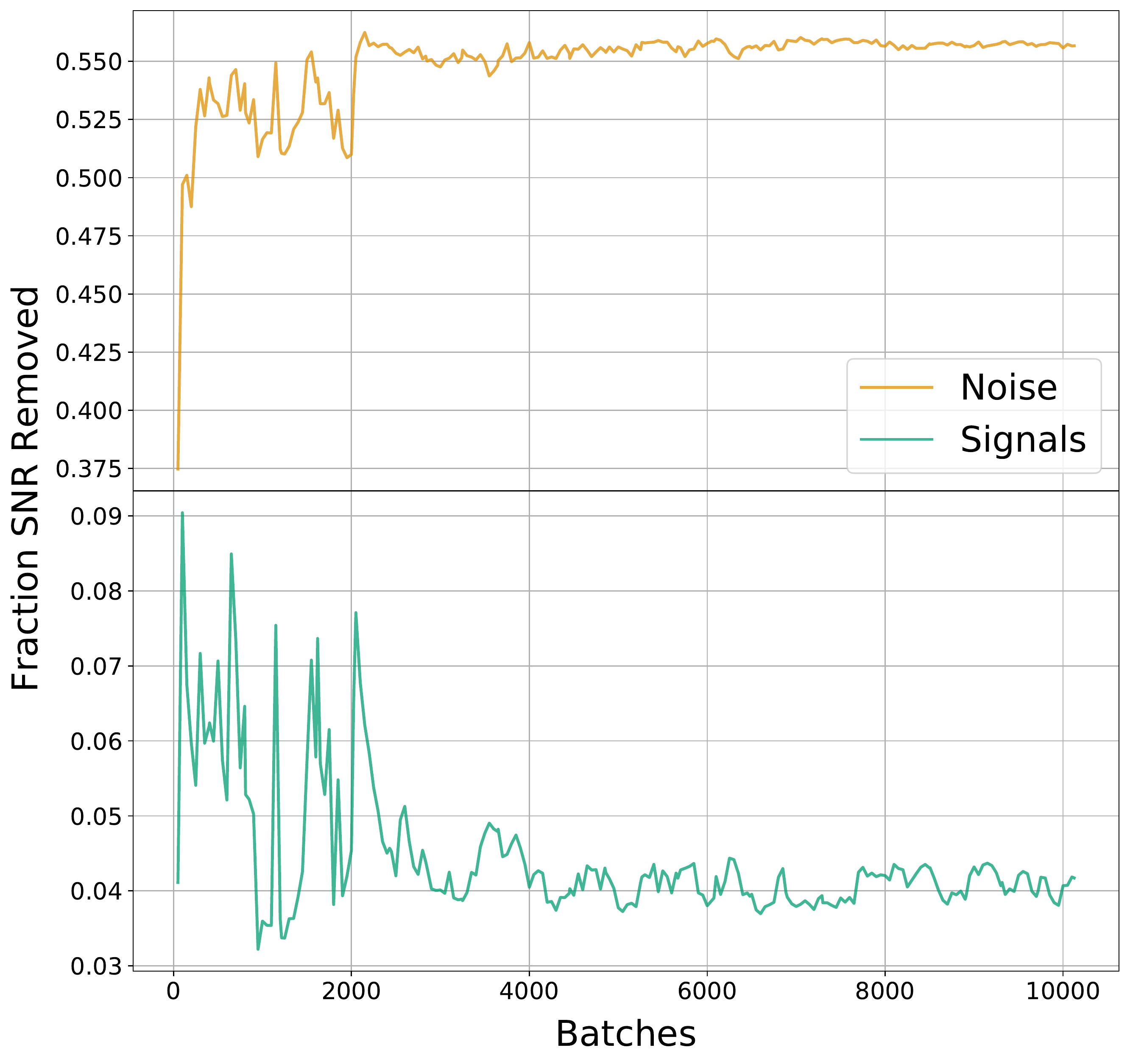}
    \caption{The average fraction of the \snr removed calculated using the unseen test data as it changes with the number of training batches used, split by noise and signal samples.}
    \label{fig:metric}
\end{figure}

As shown in Fig. \ref{fig:loss} we see that the loss calculated using the test set decreases as training continues. This change is mainly driven by the model down-weighting noise triggers, while the contribution from signals makes a smaller change over the course of the training. The effect of this training can also be seen in Fig. \ref{fig:metric}, we can see that as training continues a larger fraction of the \snr is removed as noise triggers are targeted more effectively by the model, while signals are left relatively unaffected.

\begin{figure}[t]
    \centering
    \includegraphics[width=0.85\linewidth]{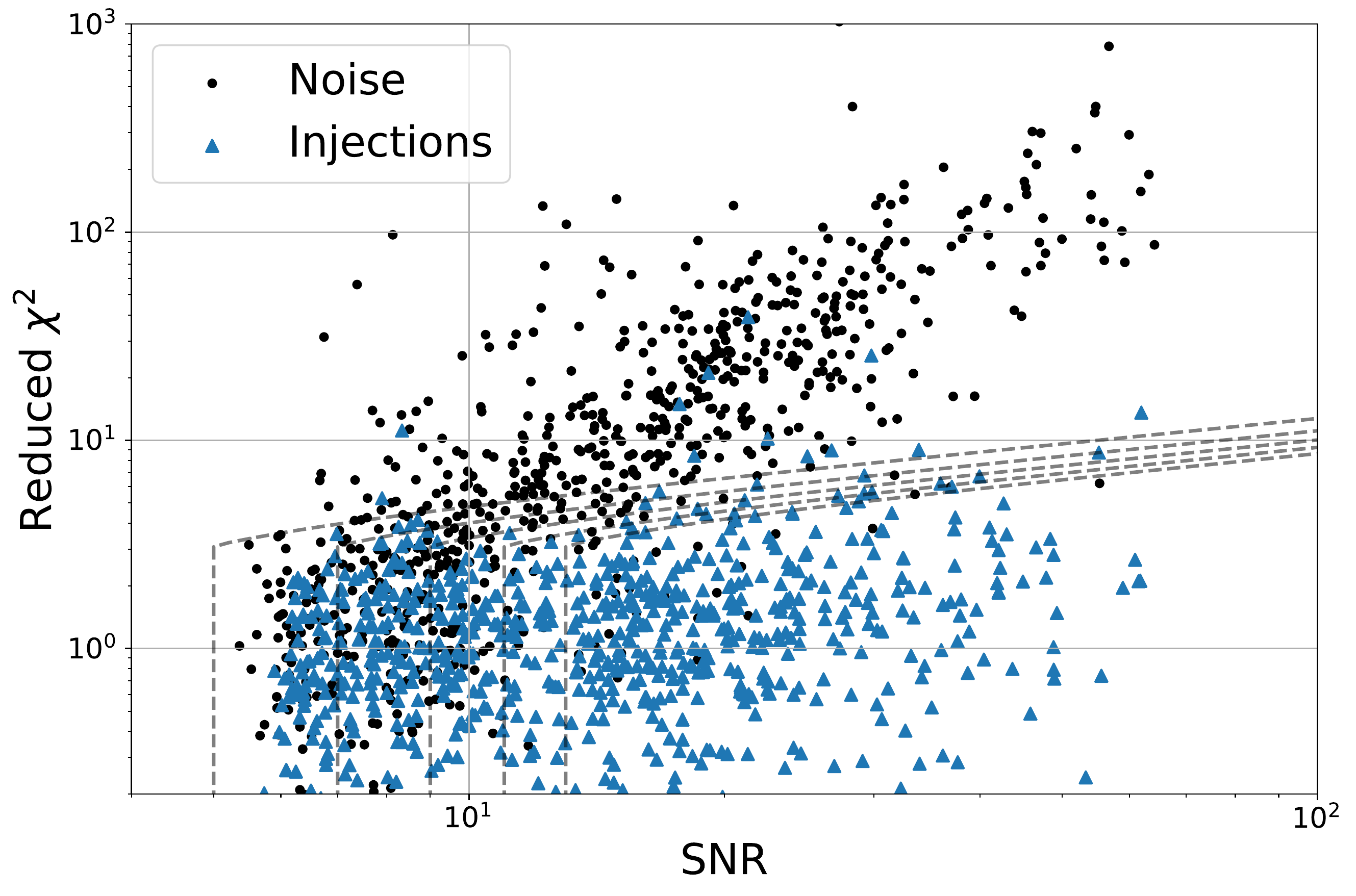}
    \caption{The test samples plotted in the \snr-\chisq plane after training is complete. Lines of constant re-weighted \snr are plotted. We can see that at high \snr values there is good separation between the noise samples (black dots) and signal samples (blue triangles), allowing the model to down-weight noise triggers heavily.}
    \label{fig:snr_chi}
\end{figure}

In Fig. \ref{fig:snr_chi} we can see that the noise and signal populations are well separated in the \snr-\chisq plane, particularly at high \snr[s]. The parameters in Eq. \ref{eq:reweight} have been trained such that the majority of signal samples are below the threshold, $\sigma$, while noise samples above the threshold are heavily down-weighted.

The trained parameters are available as a supplementary data file at: \url{https://icg-gravwaves.github.io/chisqnet/}

\section{\label{sec:results}Effect on a Intermediate Mass Black Hole Search}

In this section we will show the effect of introducing this model to a search for \imbh[s]. For this test we carry out a search following the configuration set out in \cite{pycbcimbh}, covering $~ 8$ days of data from the first half of the third observing run. We run the search twice, with the only change being the introduction of the model trained in the previous section. To ensure that the model generalises to new data we run this search on a stretch of data completely separate to that used during training.

\begin{figure}[t]
    \centering
    \includegraphics[width=0.85\linewidth]{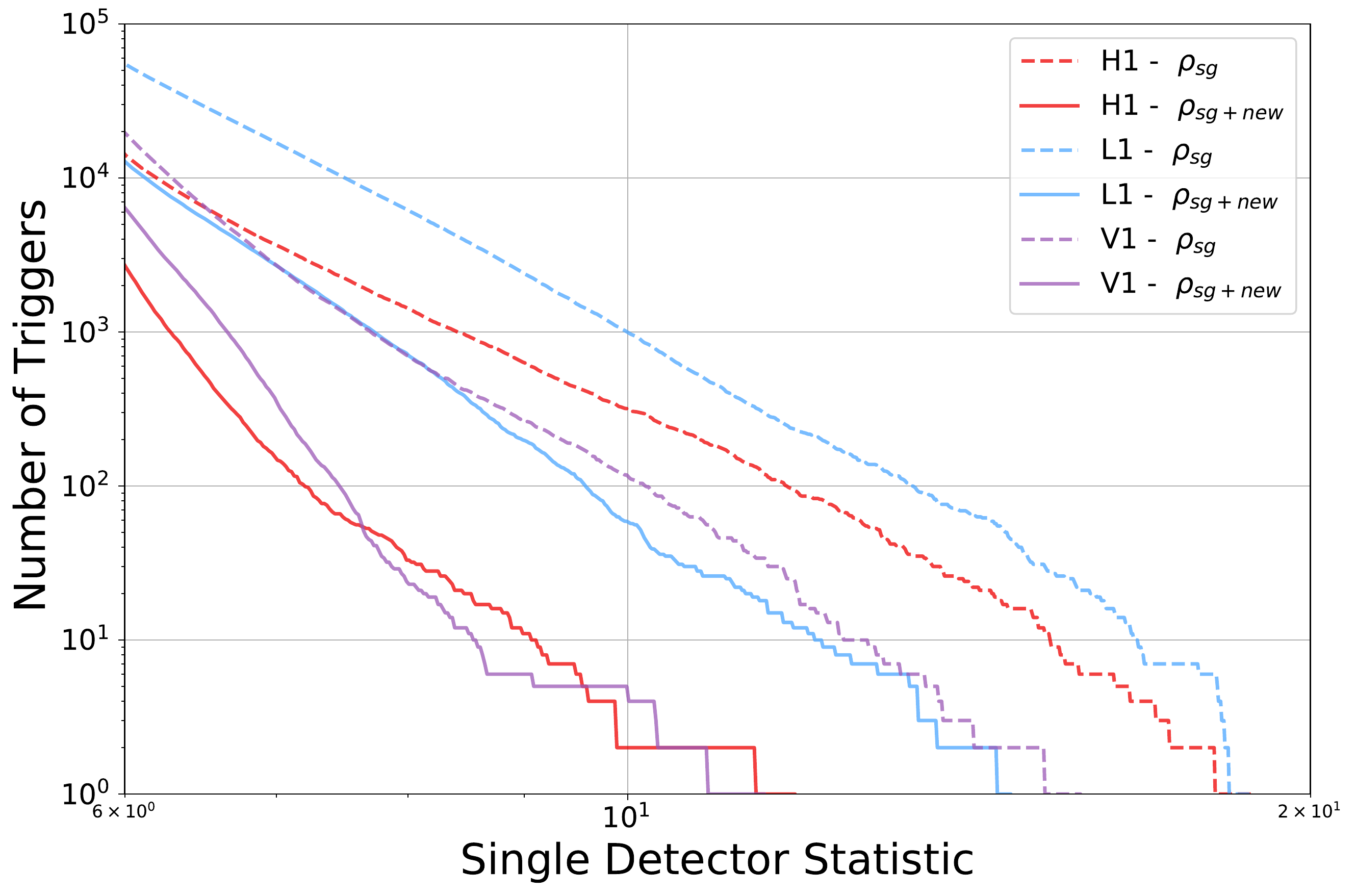}
    \caption{The cumulative number of single-detector triggers with ranking statistics below a given value. We can see a reduction in the number of single-detector triggers for all three detectors when changing from the previous ranking statistic (dashed line) to the new ranking statistic (solid line) including the new tuned model.}
    \label{fig:cumnum}
\end{figure}

By introducing this model, noise triggers can be effectively down-weighted. Fig. \ref{fig:cumnum} shows the change in the number of triggers found when using the new ranking statistic. This reduction in triggers will reduce the number of coincident noise triggers in the foreground and the empirically measured background. It is this decrease in the rate of background triggers that produces an increase in the significance of remaining foreground triggers, thereby improving the sensitivity of the search.

We evaluate the sensitivity of the search using a number of simulated signals added to the data and analysed in the same way as the main search. These simulated signals follow the same distribution as those in \cite{pycbcimbh}. The \textsc{SEOBNRv4} and \textsc{SEOBNRv4HM} waveform models are used to generate aligned spin signals, with total masses in the range $[100, 600] \msun$ and mass ratios in the range $[1, 10]$. Precessing signals are generated using the \textsc{NRSur7DQ4} waveform model with total masses in the range $[100, 600] \msun$, mass ratios in the range $[1, 4]$ and component spins isotropically distributed. For all simulated signals the distance is drawn uniformly in the range $[0.5, 11]$ Gpc, with isotropic sky positions and binary orientations. The sensitive volume of the search is then calculated by applying a threshold to the calculated false-alarm rate of 1 per year and measuring the detection efficiency in a number of distance bins. The detection efficiencies are then multiplied by the volume enclosed in the distance bins and the volumes summed to find the total sensitive volume of the search.

\begin{figure}[t]
    \centering
    \includegraphics[width=0.85\linewidth]{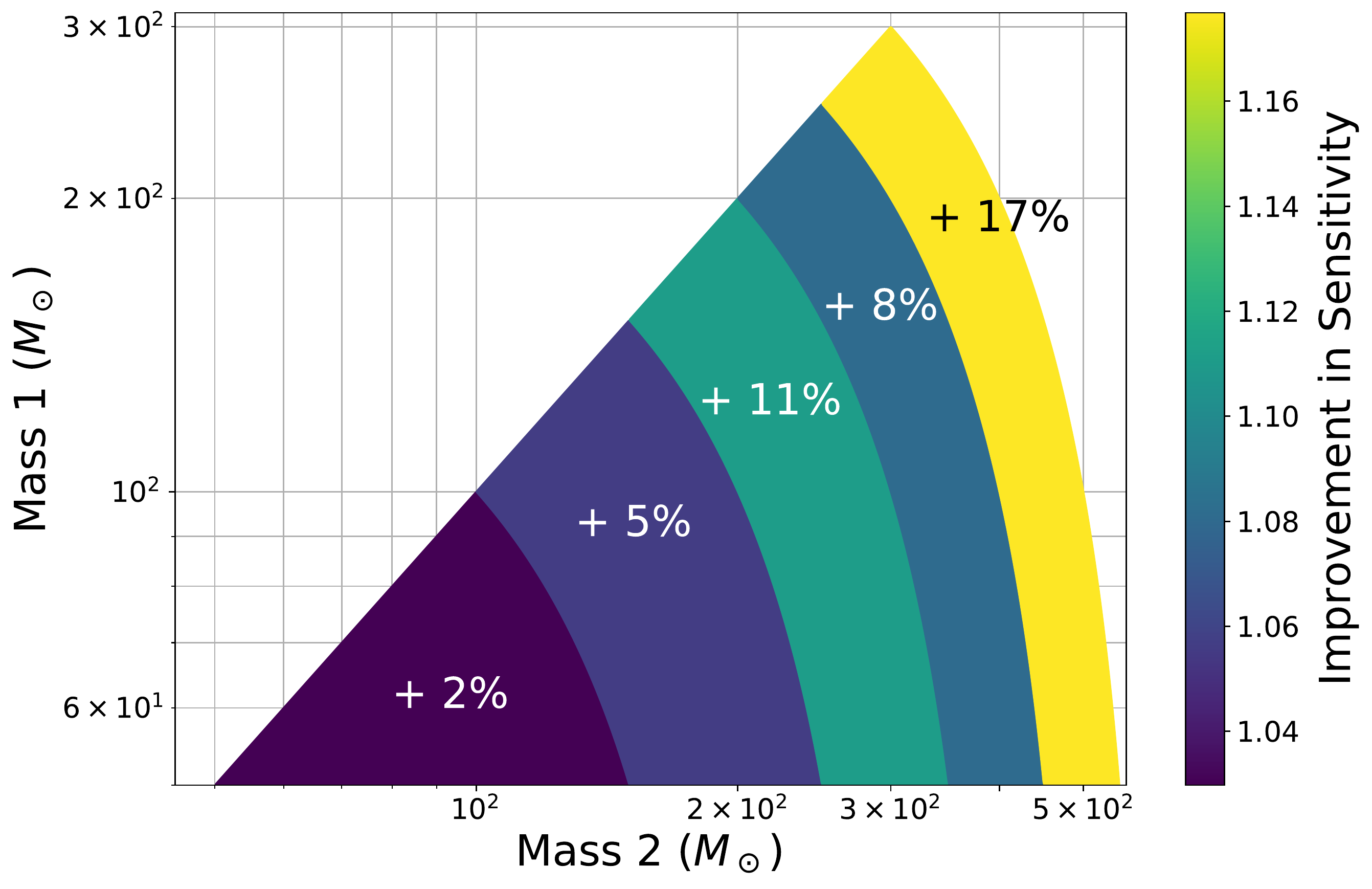}
    \caption{The ratio of the sensitive volume-time for the search including the trained model to the search without, calculated using simulated signals added to the data with a detection threshold on false-alarm rate of 1 per year.}
    \label{fig:vt}
\end{figure}

In Fig. \ref{fig:vt} we see that the sensitivity of the search has been increased by the addition of the new \chisq test, with an increase in sensitivity of $\sim 4\%$ for signals with total masses in the range $[100, 300] \msun$, increasing to $\sim 11\%$ for signals with total masses in the range $[300, 600] \msun$. This is due to the higher rate of glitches matching high mass templates, any decrease in the glitch population will therefore have a larger effect for these templates.

\section{\label{sec:discussion}Conclusion and Outlook}

In this work we have demonstrated a new framework to automatically train complex new \chisq signal-consistency tests within modelled searches for \gw signals. We have applied this to the example of a search for \imbh signals, where glitches have a strong effect on the sensitivity of the search. Our framework is able to train a new \chisq model, which provides an improved separation of the signal and noise populations allowing the noise triggers to be down-weighted. Using this new \chisq test improves the sensitivity of the search to real signals by $\sim 4\%$ for signals with total masses in the range $[100, 300] \msun$ and $\sim 11\%$ for signals with total masses in the range $[300, 600] \msun$

The introduction of new \chisq tests is difficult and usually requires empirical tuning by hand to be effective, and often requires re-tuning for different target parameter spaces or noise populations. As signal-consistency tests become more complex this can become unfeasible. However, by utilising machine-learning techniques we have shown that we can tune these automatically, removing the burden in improving and optimally tuning these tests. The method we demonstrate here could be applied to any of the commonly used matched-filter search pipelines targeting compact binary mergers.

The population of glitches within the interferometer data continues to be one of the largest challenges facing \gw searches. By continuing to develop signal-consistency tests that specifically target such noise we can continue to improve the sensitivity of searches and increase the chance of observing new events in areas of the parameter space most affected by glitches.

\begin{acknowledgments}
CM was supported by the Science and Technology Facilities Council through the DISCnet Centre for Doctoral Training.
IH was supported by STFC grants ST/T000333/1 and ST/V005715/1.

This research has made use of data or software obtained from the Gravitational Wave Open Science Center (gw-openscience.org), a service of LIGO Laboratory, the LIGO Scientific Collaboration, the Virgo Collaboration, and KAGRA. LIGO Laboratory and Advanced LIGO are funded by the United States National Science Foundation (NSF) as well as the Science and Technology Facilities Council (STFC) of the United Kingdom, the Max-Planck-Society (MPS), and the State of Niedersachsen/Germany for support of the construction of Advanced LIGO and construction and operation of the GEO600 detector. Additional support for Advanced LIGO was provided by the Australian Research Council. Virgo is funded, through the European Gravitational Observatory (EGO), by the French Centre National de Recherche Scientifique (CNRS), the Italian Istituto Nazionale di Fisica Nucleare (INFN) and the Dutch Nikhef, with contributions by institutions from Belgium, Germany, Greece, Hungary, Ireland, Japan, Monaco, Poland, Portugal, Spain. The construction and operation of KAGRA are funded by Ministry of Education, Culture, Sports, Science and Technology (MEXT), and Japan Society for the Promotion of Science (JSPS), National Research Foundation (NRF) and Ministry of Science and ICT (MSIT) in Korea, Academia Sinica (AS) and the Ministry of Science and Technology (MoST) in Taiwan.

For the purpose of open access, the author has applied a CC BY public copyright licence to any Author Accepted Manuscript version arising
\end{acknowledgments}

\bibliography{auto_tuned_chisq}

\appendix

\end{document}